\documentclass[12pt,sort&compress]{elsarticle}
\pdfoutput=1

\usepackage[utf8]{inputenc}

\usepackage{graphics}
\usepackage{graphicx}
\usepackage{amssymb}
\usepackage{amsthm}  

\usepackage[nolist,nohyperlinks]{acronym}
\usepackage{glossaries}
\usepackage{hyperref}
\usepackage{caption}
\usepackage{subcaption}
\usepackage{interval}
\usepackage{stmaryrd}
\usepackage{hyphenat}
\usepackage{dirtytalk}
\usepackage{amsmath}
\usepackage{booktabs}
\usepackage{cancel}
\usepackage{cleveref}
\usepackage{colortbl}
\usepackage{csquotes}
\usepackage{helvet}
\usepackage{mathpazo}
\usepackage{multirow}
\usepackage{listings}
\usepackage{pgfplots}
\usepackage{xcolor}
\usepackage{siunitx}
\usepackage[normalem]{ulem}
\usepackage{scrextend}
\usepackage[export]{adjustbox}
\usepackage[T1]{fontenc}
\usepackage{tikz}

\useunder{\uline}{\ul}{}

\newcounter{spcounter}\renewcommand{\thespcounter}{\alph{spcounter}}
\newcommand{\vollabel}[1]{%
\refstepcounter{spcounter}\textsuperscript{(\thespcounter)}\label{#1}%
}

\newcommand{\Tstrut}{\rule{0pt}{2.6ex}}         
\newcommand{\Bstrut}{\rule[-0.9ex]{0pt}{0pt}}   

\newtheorem{definition}{Def.}

\newcommand{\code}[1]{\texttt{#1}}

\newcommand{\sequenceset}[1]{\llbracket #1 \rrbracket}

\newcommand{\datavol}{x}
\newcommand{\segvol}{y}
\newcommand{\gtvol}{y}

\newcommand{\probavol}{\hat{\gtvol}}

\newcommand{\vox}[2]{{#1}_{#2}}

\newcommand{\gtvox}[1]{\vox{\gtvol}{#1}}

\newcommand{\probavox}[1]{\vox{\probavol}{#1}}

\newcommand{\widthh}{w}  
\newcommand{\height}{h}
\newcommand{\depth}{d}
\newcommand{\nchannels}{c}

\newcommand{\datavoxdom}{\interval{0}{1}}
\newcommand{\datavoldom}{\mathcal{X}}
\newcommand{\datavoldomdef}{\datavoxdom^{\widthh \times \height \times \depth}}

\newcommand{\nclasses}{C}

\newcommand{\segvoxdom}{\sequenceset{\nclasses}}
\newcommand{\segvoldom}{\mathcal{Y}}
\newcommand{\segvoldomdef}{\segvoxdom^{\widthh \times \height \times \depth}}

\newcommand{\fzero}{f_0}

\newcommand*{\newterm}[2][]{\newglossaryentry{#2}{name={#2},description={\nopostdesc},#1}}

\newterm[name={Polyamide 66}]{poly}
\newterm[name={glass fiber-reinforced \gls{poly}}]{gf-poly}
\newterm[name={Deep Learning}]{dl}
\newterm[name={NVIDIA}]{nvidia}
\newterm[name={Quadro P4000}]{qpfourk}
\newterm[name={Quadro P2000}]{qptwok}
\newterm[name={CUDA}]{cuda}
\newterm[name={\code{tensorflow-gpu}}]{tf-gpu}
\newterm[name={Non-Local Means}]{non-local-means}
\newterm[name={Synchrotron SOLEIL}]{soleil}
\newterm[name={X-ray}]{xray}
\newterm[name={Psiché}]{psyche}
\newterm[name={LuAG}]{luag}
\newterm[name={Hamamatsu}]{hamamatsu}
\newterm[name={CMOS}]{cmos}
\newterm[name={PyHST2}]{pyhsttwo}
\newterm[name={VGStudioMax}]{vgstudiomax}
\newterm[name={U-depth}]{udepth}
\newterm[name={U-level}, ]{ulevel}

\newterm[name={\textit{Sister}}]{sister}
\newterm[name={\textit{Bending}}]{bending}
\newterm[name={\textit{crack}}]{biax}
\newterm[name={\textit{Train-Val-Test}}]{train-val-test}

\newterm[name={$Jaccard^2$}]{jacc2}
\newterm[name={Jaccard index},plural={Jaccard indices}]{jaccidx}
\newterm[name={Adam}]{adam}
\newterm[name={AdaBelief}]{adabelief}

\newterm[name={Fiji}]{fiji}
\newterm[name={ImageJ}]{imagej}
\newterm[name={Avizo}]{avizo}
\newterm[name={TensorFlow}]{tf}
\newterm[name={Keras}]{keras}
\newterm[name={PyPi}]{pypi}
\newterm[name={GitHub}]{gh}

\newterm[name={Pymicro}]{pymicro}
\newterm[name={BIGMÉCA}]{bigmeca}
\newterm[name={\code{tomo2seg}}]{tts}  
\newterm[name={MINES ParisTech}]{mines}

\newterm[name={Modular U-Net}]{modunet}
\newterm[name={Modular U-Net++}]{modunetpp}

\newterm[name={SegNet}]{segnet}
\newterm[name={U-Net}]{unet}

\newterm[name={ReLU}]{relu}
\newterm[name={max pooling}]{maxpool}
\newterm[name={up-sampling}]{upsampling}
\newterm[name={batch normalization}]{batchnorm}
\newterm[name={transposed convolution}]{transpconv}
\newterm[name={Zenodo}]{zenodo}
\newterm[name={layer normalization}]{layernorm}

\begin{acronym}[FCNN]

\acro{xct}[XCT]{X-ray Computed Tomography}
\acro{srg}[SRG]{Seeded Region Growing}
\acro{cor}[CoR]{Center of Rotation}
\acro{lom}[LOM]{Light Optical Microscopy}
\acro{sem}[SEM]{Scanning Electron Microscopy}
\acro{ldct}[LDCT]{Low-Dose \ac{xct}}
\acro{ndct}[NDCT]{Normal-Dose \ac{xct}}
\acro{mlem}[MLEM]{likelihood expectation maximization algorithm}
\acro{ss}[SS]{Serial Sectioning}
\acro{iou}[IoU]{Intersection over Union}
\acro{sgd}[SGD]{Stochastic Gradient Descent}
\acro{lr}[LR]{Learning Rate}
\acro{gpu}[GPU]{Graphics Processing Unit}
\acro{roc}[ROC]{Receiver Operating Characteristic}
\acro{mdf}[MDF]{The Materials Data Facility}
\acro{esrf}[ESRF]{European Synchrotron Radiation Facility}

\acro{gan}[GAN]{Generative Adversarial Network}
\acrodefplural{gan}{Generative Adversarial Network}

\acro{nn}[NN]{Neural Network}
\acrodefplural{NN}{Neural Networks}

\acro{dnn}[DNN]{Deep \ac{nn}}
\acrodefplural{DNN}{Deep \acp{nn}}

\acro{cnn}[CNN]{Convolutional \ac{nn}}
\acrodefplural{CNN}{Convolutional \acp{nn}}

\acro{fcnn}[FCNN]{Fully-convolutional \ac{nn}}
\acrodefplural{FCNN}{Fully-convolutional \acp{nn}}

\acro{convblock}[ConvBlock]{\textit{Convolutional Block}}
\acro{upsample}[UpBlock]{\textit{Upsampling Block}}
\acro{downsample}[DownBlock]{\textit{Downsampling Block}}

\end{acronym}

\definecolor{segm:blue}{RGB}{75, 77, 239}
\definecolor{segm:yellow}{RGB}{220, 147, 15}
\definecolor{segm:red}{RGB}{211, 0, 21}

\newcommand{\linkTts}{\href{https://github.com/joaopcbertoldo/tomo2seg}{\nolinkurl{github.com/joaopcbertoldo/tomo2seg}}}

\newcommand{\linkAvizo}{\href{https://www.thermofisher.com/fr/fr/home/industrial/electron-microscopy/electron-microscopy-instruments-workflow-solutions/3d-visualization-analysis-software/avizo-materials-science.html}{\nolinkurl{thermofisher.com}}}

\newcommand{\linkVgstudiomax}{\href{https://www.volumegraphics.com/en/products/vgstudio-max.html}{\nolinkurl{volumegraphics.com/en/products/vgstudio-max.html}}}

\newcommand{\linkPypiTfGpu}{https://pypi.org/project/tensorflow-gpu/}

\newcommand{\linkBigmeca}{\url{https://bigmeca.minesparis.psl.eu/}}
\newcommand{\linkSoleil}{\url{https://www.synchrotron-soleil.fr/fr}}

\newcommand{\linkNvidiaQuadroPFourK}{\href{https://www.pny.com/nvidia-quadro-p4000}{\nolinkurl{pny.com/nvidia-quadro-p4000}}}

\newcommand{\linkNvidiaQuadroPTwoK}{\href{https://www.pny.com/nvidia-quadro-p2000}{\nolinkurl{pny.com/nvidia-quadro-p2000}}}

\newcommand{\linkOpenVolumesDemo}[1]{\href{https://github.com/joaopcbertoldo/gfpa66-volumes}{#1}}

\newcommand{\linkYoutubeBiaxSurface}{\href{https://youtu.be/rmBTZrcMrCk}{\nolinkurl{youtu.be/rmBTZrcMrCk}}}
\newcommand{\linkYoutubeTestVolSegm}{\href{https://youtu.be/HvdWhDZJgLE}{\nolinkurl{youtu.be/HvdWhDZJgLE}}}
\newcommand{\linkYoutubeTestVolErr}{\href{https://youtu.be/dXlYcLXHFAA}{\nolinkurl{youtu.be/dXlYcLXHFAA}}}
\newcommand{\linkYoutubeTestVolSegmTwo}{\href{https://youtu.be/CjwG-1FoSCY}{\nolinkurl{youtu.be/CjwG-1FoSCY}}}
\newcommand{\linkYoutubeRawData}{\href{https://youtu.be/4kifxlvxzb8}{\nolinkurl{youtu.be/4kifxlvxzb8}}}
\newglossary*{tm}{Theoretical models}
\newglossary*{volfiles}{Volume files}

\journal{Nature Machine Intelligence}

\begin{document}

\begin{frontmatter}
\title{
A modular \gls{unet} for automated segmentation of \gls{xray} tomography images in composite materials
}

\author[cdm]{
João P C Bertoldo
\fnref{joao}
}

\author[cmm]{
Etienne Decencière
\fnref{etienne}
}

\author[cdm]{
David Ryckelynck
\fnref{david}
}

\author[cdm]{
Henry Proudhon
\fnref{henry}
}

\newcommand{\orcid}[1]{%
\includegraphics[width=9pt]{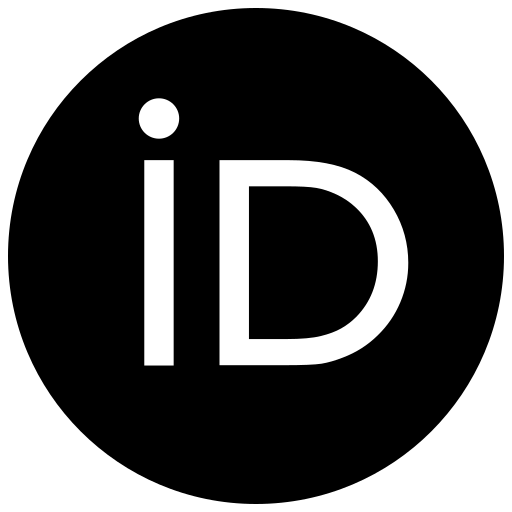}%
\hspace{2pt}%
\href{https://orcid.org/#1}{\texttt{#1}}%
}

\fntext[joao]{\orcid{0000-0002-9512-772X}}
\fntext[etienne]{\orcid{0000-0002-1349-8042}}
\fntext[david]{\orcid{0000-0003-3268-4892}}
\fntext[henry]{\orcid{0000-0002-4075-5577}}

\address[cdm]{Centre des Matériaux}
\address[cmm]{Centre de Morphologie Mathématique}

\address{MINES ParisTech, PSL Research University, France}

\date{July 2021}

\begin{abstract}
\ac{xct} techniques have evolved to a point that high-resolution data can be acquired so fast that classic segmentation methods are prohibitively cumbersome, demanding automated data pipelines capable of dealing with non-trivial 3D images.
Deep learning has demonstrated success in many image processing tasks, including material science applications, showing a promising alternative for a human-free segmentation pipeline.
In this paper a modular interpretation of \gls{unet} (\gls{modunet}) is proposed and trained to segment 3D tomography images of a three-phased \gls{gf-poly}.
We compare 2D and 3D versions of our model, finding that the former is slightly better than the latter.
We observe that human-comparable results can be achievied even with only 10 annotated layers and using a shallow \gls{unet} yields better results than a deeper one.
As a consequence, \ac{nn} show indeed a promising venue to automate \ac{xct} data processing pipelines needing no human, adhoc intervention.
\end{abstract}

\begin{keyword}
Deep Learning \sep
U-Net  \sep
Modular Network Architecture \sep
Semantic Segmentation \sep
3D X-ray Computed Tomography \sep
Composite Material
\end{keyword}
\end{frontmatter}

\setcounter{footnote}{1}

%

\section*{}
\label{sec:intro}
\acresetall

\ac{xct}, a characterization technique used by material scientists for non-invasive analysis, has tremendously progressed over the last 10 years with improvements in both spatial resolution and throughput~\cite{withersXrayComputedTomography2021,Maire2014}.
Progress with synchrotron sources, including the recent \ac{esrf} upgrade~\cite{Pacchioni_2019}, made it possible to look inside a specimen without destroying it in a matter of seconds~\cite{Shuai_AM_2016} -- sometimes even faster~\cite{Maire_IJFR_2016}.

This results in a wealth of 3D tomography images (stack of 2D images) that need to be analyzed and, in some applications, it is desirable to segment them (i.e.\ transform the gray-scaled voxels in semantic categorical values).
A segmented image is crucial for quantitative analyses;
for instance, measuring the distribution of precipitate length and orientation~\cite{kaira-auto-seg-2018}, or phase characteristics, which can be useful for more downstream applications like estimating thermo-mechanical properties~\cite{semantic-seg-tomo-2019}.

\ac{xct} images typically have billions of voxels, weighting several gigabytes, and remain complex to inspect manually even using dedicated costly software (e.g.: \gls{avizo}\footnote{\linkAvizo}, \gls{vgstudiomax}\footnote{\linkVgstudiomax}).
Using thresholds on the gray level image is an easy, useful method to segment phases in tomographies, but it fails in complex cases, in particular when acquisition artifacts (e.g.: rings, beam hardening, phantom gradients) are present.
Algorithms based on mathematical morphology like the watershed segmentation~\cite{beucherWatershed, beucherWatershed2} help tackling more complex scenarios, but they need human parametrization, which often requires expertise in the application domain.
Thus, scaling quantitative analyses is expensive, creating a bottleneck to process 3D \ac{xct} -- or even 4D (3D with time steps).

\gls{dl} approaches offer a viable solution to attack this issue because \acp{nn} can generalize patterns learned from annotated data.
A \ac{nn} is a statistical model originated from perceptrons~\cite{perceptron} capable of approximating a generic function. \acp{cnn}~\cite{cnn-bengio-lecun}, a variation adapted to spatially-structured data (time series, images, volumes), made great advances in computer vision tasks.
Since the emergence of popular frameworks like \gls{tf}~\cite{tensorflow2015}, more problem-specific architectures have been proposed, such as Fully-convolutional \acp{nn}~\cite{fcnn}, a convolution-only type of model used to map image pixel values to another domain (e.g. classification or regression).

\cite{kaira-auto-seg-2018} trained a model to segment three phases in 3D nanotomographies of an Al-Cu alloy, showing that even a simple \ac{cnn} can reproduce patterns of a human-made segmentation.
\cite{optim-cnn-stan-2020} optimized a \gls{segnet}~\cite{segnet} to segment dendrites of different alloys, including a 4D \ac{xct}.
\cite{semantic-seg-tomo-2019} identified Aluminides and Si phases in \ac{xct} using a \gls{unet}, an architecture that, along with its many flavors~\cite{unet, unet3d, unetsquared, unetpp}, has shown success in a variety of applications~\cite{oktayAttentionUNetLearning2018, stollerWaveUNetMultiScaleNeural2018, zhangRoadExtractionDeep2018}.
Finally, \cite{furatMachineLearningTechniques2019} combined \glspl{unet} with classic segmentation algorithms (e.g.\ marker-based watershed) to segment grain boundaries in successive \acp{xct} of an Al-Cu specimen as it is submitted to Ostwald ripening steps.

In this paper, an annotated 3D \ac{xct} of \gls{gf-poly} is presented as an example of segmentation problem in Materials Science that can be automated with a deep learning approach.
Our \ac{nn} architecture, the \gls{modunet} (Fig.~\ref{fig:modunet} and Fig.~\ref{fig:modunet-blocks}), is proposed as a generalized representation of the \gls{unet}, explicitly factorizing the U-like structure from its composing blocks.

Like~\cite{furatMachineLearningTechniques2019}, we compare three variants on the composite material dataset focusing on the dimensionality of the convolutions (2D or 3D), obtaining qualitatively human-like segmentation (Fig.~\ref{fig:test-segm-zooms} and Fig.~\ref{fig:biax}) with all of them although 2D-convolutions yield better results (Fig.~\ref{fig:models-perf-chart-modelvars}).
We find that (for the considered material) the \gls{unet} architecture can be shallow without loss of performance, but batch normalization is necessary for the optimization (Fig.~\ref{fig:models-perf-chart-modelstripping}).
Finally, a model's learning curve (Fig.~\ref{fig:learning-curve}) shows that only ten annotated 2D slices are necessary to train our \ac{nn}.

Our results show that \acp{nn} can be not only a quality-wise satisfactory but also a viable solution in practice for \ac{xct} segmentation as it requires relatively little annotated data and shallow models (therefore faster to train).


\section{Data}
\label{sec:data}

\begin{figure}[!t]

\captionsetup{font=scriptsize, labelfont=scriptsize, width=.9\textwidth}
\centering

\begin{subfigure}[b]{.48\textwidth}

\centering
\captionsetup{font=scriptsize, labelfont=scriptsize, width=.9\textwidth}
\includegraphics[height=61mm]{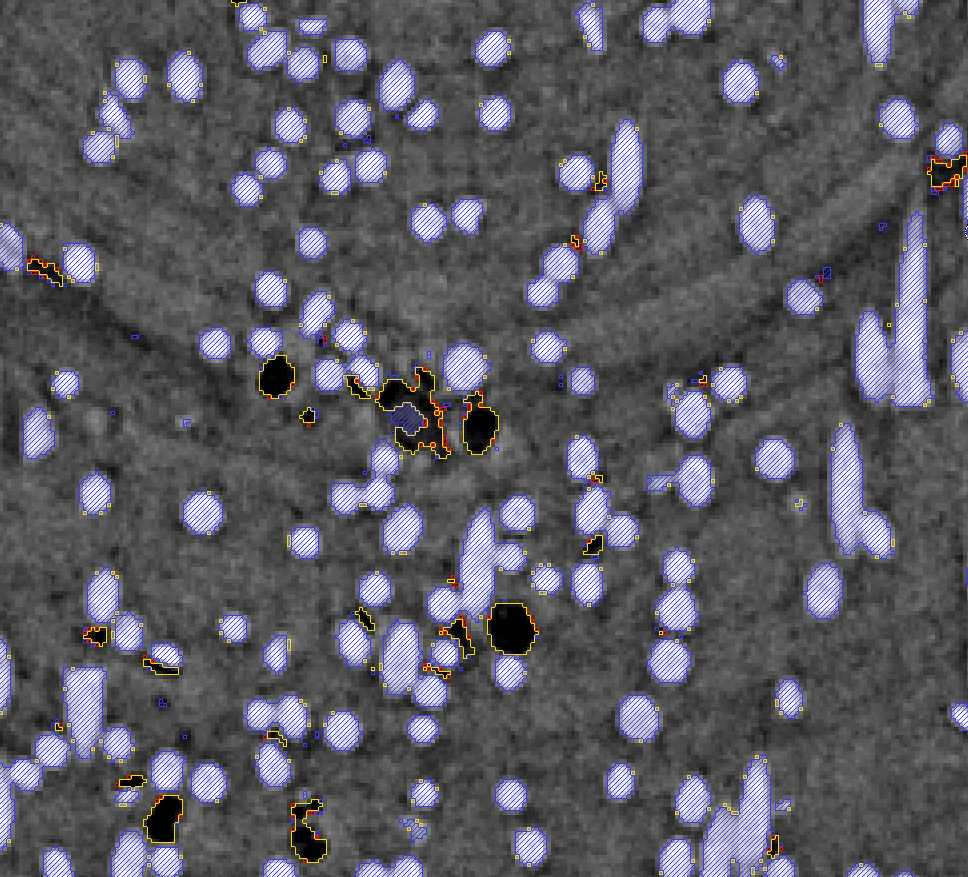}

\end{subfigure}
\hspace{0.8mm}
\begin{subfigure}[b]{.48\textwidth}

\centering
\captionsetup{font=scriptsize, labelfont=scriptsize, width=.9\textwidth}
\includegraphics[height=61mm]{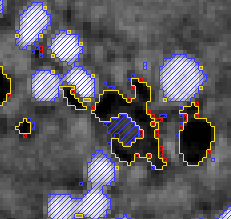}

\end{subfigure}

\caption{
Segmentation on the test set.
Color code: \textcolor{segm:blue}{blue represents voxels \textit{correctly} classified as fiber (hatched)}, \textcolor{segm:yellow}{yellow as porosity (contours)}, and \textcolor{segm:red}{red represents misclassification}.
Supplementary video: \linkYoutubeTestVolSegm.
}
\label{fig:test-segm-zooms}

\end{figure}

The data used in this work is composed of synchrotron \gls{xray} tomography volumes recorded using \SI{2}{\milli\metre} $\times$ \SI{2}{\milli\metre} cross section composite specimens of \gls{poly} reinforced by glass fibers.
A volume of $2048^3$ voxels, referred to as \gls{train-val-test} (Fig.~\ref{fig:test-segm-zooms} and Fig.~\ref{fig:raw-data}), was cropped to get rid of the specimen's borders, and its ground truth segmentation was created semi-manually with \gls{imagej}~\cite{imagej} (using \gls{fiji}~\cite{fiji}) and \gls{avizo}.
The tomography of another specimen of the same material was also processed and partially annotated in order to evaluate our models (Fig.~\ref{fig:biax}) -- it is further referred to as the \say{\gls{biax}} volume because of the fracture inside it.
Both image volumes and the former's annotations are publicly available~\cite{zenodoData}.

\paragraph{Acquisition}

\gls{xray} tomography scans were recorded on the \gls{psyche} beamline at the \gls{soleil} using a parallel pink beam.
The incident beam spectrum was characterized by a peak intensity at \SI{25}{\kilo\eV}, defined by the silver absorption edge, with a full width at half maximum bandwidth of approximately \SI{1.8}{\kilo\eV}.
The total flux at the sample position was about \SI{2.8e12}{photons/s/mm^2}.
The detector placed after the sample was constituted by a \gls{luag} scintillator, a 5$\times$ magnifying optics, and a \gls{hamamatsu} \gls{cmos} 2048 x 2048 pixels detector (effective pixel size of \SI{1.3}{\micro\metre}).
1500 radiographs were collected over a \ang{180} rotation and an exposure of \SI{50}{ms} (full scan duration of 2 minutes).
The sets of radiographs were then processed using \gls{pyhsttwo} reconstruction software~\cite{Mirone_Nimb_2014} with the Paganin filter~\cite{Paganin_JM_2002} activated to enhance the contrast between the phases.

\paragraph{Phases}

The three phases present in the material are visible in Fig.~\ref{fig:test-segm-zooms}: the polymer matrix (gray), the glass fibers (white, hatched in blue), and damage in form of pores (dark gray and black, contoured in yellow) -- referred as \textit{porosity} here.
One can observe that the orientation of the fibers is unevenly distributed;
they are mostly along the axes Y (vertical) and Z (out of the plane) in Fig.~\ref{fig:raw-data}.

\paragraph{Ground truth}
The data was annotated in two steps: first, the fiber and the porosity phases were independently segmented using \acl{srg}~\cite{srg};
then, ring artifacts that leaked to the porosity class were manually corrected.
A detailed description of the procedure is presented in the~\ref{sec:data-further}.

\paragraph{Data split}

The ground truth layers (of the \gls{train-val-test} volume) were sequentially split -- their order were preserved to train the 3D models (Section~\ref{subsec:modelvars}) -- into three sets: train (1300 layers), validation (128 layers), and test (300 layers).
A margin of 86 layers between these sets was adopted to avoid information leakage.
The train layers were used to train the \acp{nn}, the validation layers were used to select the best model (during the optimization), and the test layers were used to evaluate the models (Section~\ref{sec:results}).

\paragraph{Class imbalance}

Due to the material's nature, the classes (phases) in this dataset are intrinsically imbalanced.
The matrix, the fiber, and the porosity represent, respectively, 82.3\%, 17.2\%, and 0.5\% of the voxels.


\section{\acl{nn}}\label{sec:nn}
\acresetall

\paragraph{Problem formulation}

Let $ \: \datavol \in \datavoldom = \datavoldomdef \: $ be a normalized gray 3D image.
Its segmentation $ \: \segvol \in \segvoldom = \segvoldomdef \: $, where $ \: \sequenceset{C} = \{0, 1, \dots, C - 1\} \: $, contains a class value in each voxel, which may represent any categorical information, such as the phase of the material.
In this setting, a segmentation algorithm is a function $ \: f: \datavoldom \to \segvoldom \: $.
In this section we present our approach (that is, the $ \: f \: $) used to segment the data described in the previous section.

First, a generic \gls{unet} architecture, which we coined \textit{\gls{modunet}} (Fig.~\ref{fig:modunet}), is proposed;
then, the modules used in this work are briefly exposed (detailed description and hyperparameters in the\ \ref{sec:default-hyperparams});
finally, three variations of the \gls{modunet}, based on the input, convolution, and output nature (2D or 3D), are presented.
Our training setup (loss function, optimizer, learning rate, data augmentation, implementation framework, and hardware) is described in\ \ref{sec:training}.

\subsection{\gls{modunet}}\label{subsec:modunet}

\begin{figure}[!t]

\captionsetup{font=scriptsize, labelfont=scriptsize, width=.9\textwidth}
\centering

\includegraphics[height=55mm]{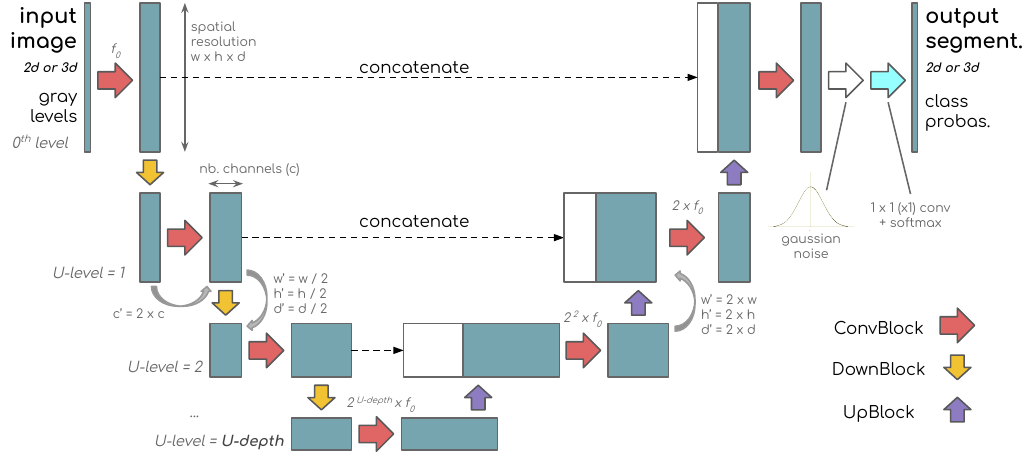}

\caption{
\gls{modunet}: a generalization of the \gls{unet} architecture.
}
\label{fig:modunet}

\end{figure}

Since \cite{unet} proposed \gls{unet}, variations of it emerged in the literature (e.g.: \cite{unet3d, unetsquared}).
Here we propose a generalized version, preserving its overall structure.
The \textit{\gls{modunet}} (Fig.~\ref{fig:modunet}) is based on three blocks: the \ac{convblock}, the \ac{downsample}, and the \ac{upsample}.

The left/right side of the architecture corresponds to an encoder/decoder, a repetition of pairs of \ac{convblock} and \ac{downsample}/\ac{upsample} modules.
They are connected by concatenations between their respective parts at the same \textit{\gls{ulevel}} -- which corresponds to the inner tensors' resolutions (higher \gls{ulevel} means lower resolution).
The \textit{\gls{udepth}}, a hyperparameter, is the number of \gls{ulevel}s in a model, corresponding to the number of \ac{downsample} (and equivalently \ac{upsample}) modules.

The \ac{convblock} is a combination of operations that outputs a tensor with the same spatial dimensions of its input, though the number of channels may differ -- in our models it always doubles.
The assumption of equally-sized input/output is optional, but we admit it for the sake of simplicity because it makes the model easier to be used with an arbitrarily shaped volume.
The numbers of channels after the \acp{convblock} is $ \: 2^{\text{\gls{ulevel}}} \times \fzero \: $, where $ \: \fzero \: $ is the number of filters in the first convolution.

The \ac{downsample}/\ac{upsample} divides/multiplies the input tensor's shape by two in every spatial dimension: width, length, and depth in the 3D case.
In other words, a tensor with shape $ \: (\widthh, \height, \depth, \nchannels) \: $, where $ \: \nchannels \: $ is the number of channels, becomes, respectively, $ \: (\frac{\widthh}{2}, \frac{\height}{2}, \frac{\depth}{2}, \nchannels) \: $ after a \ac{downsample} and $ \: (2\widthh, 2\height, 2\depth, \nchannels) \: $ after an \ac{upsample}.

In~\cite{unet}, for instance, the \ac{convblock} is a sequence of two 3x3 convolutions with \gls{relu} activation, the \ac{downsample} is a \gls{maxpool}, and the \ac{upsample} is an \gls{upsampling} layer.
In~\cite{unet3d}, the \ac{convblock} is a 3D convolutional layer, and in~\cite{unetsquared} it is a nested \gls{unet}.

\begin{figure}[!t]

\captionsetup{font=scriptsize, labelfont=scriptsize, width=.9\textwidth}
\centering

\includegraphics[width=\textwidth]{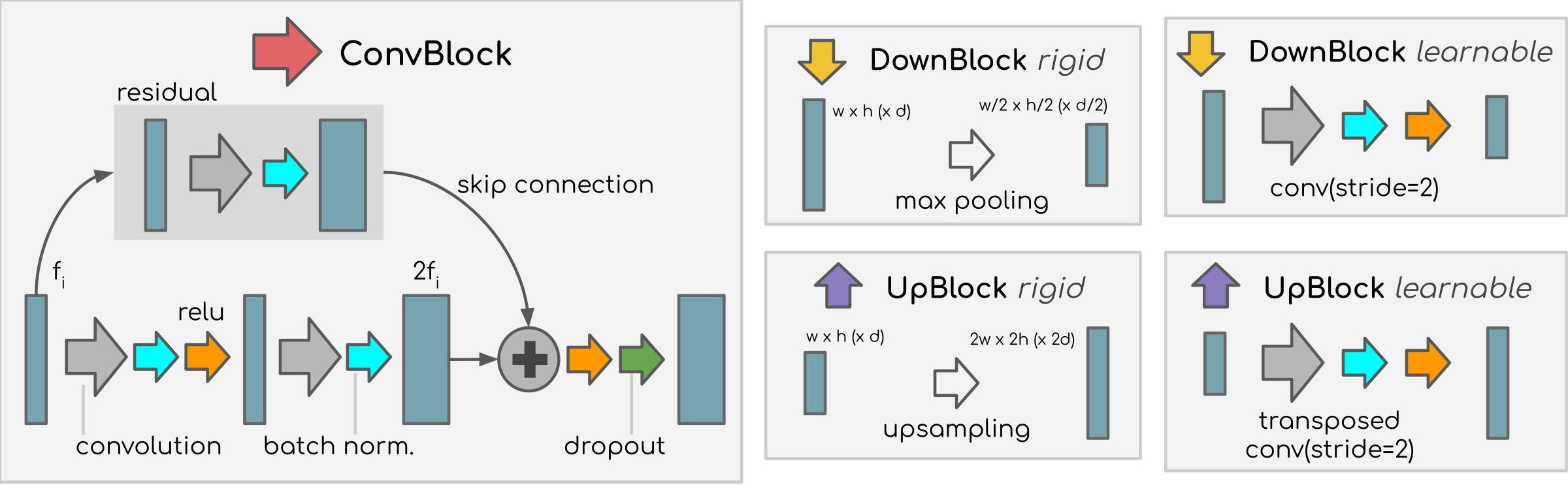}

\caption{
Examples of \gls{modunet} blocks.
Left: our \ac{convblock}.
Middle/right: rigid/learnable \ac{downsample} and \ac{upsample}.
}

\label{fig:modunet-blocks}

\end{figure}

The \ac{convblock} used here (Fig.~\ref{fig:modunet-blocks}) is a sequence of two 3x3 (x3 in the 3D case) convolutions with \gls{relu} activation, a residual connection with another convolution, batch normalization before each activation, and dropout at the end.
The \ac{downsample} is a $3 \times 3$ convolution with $2 \times 2$ stride, and the \ac{upsample} is a $3 \times 3$ \gls{transpconv} with $2 \times 2$ stride.

\subsection{Variations: 2D, 2.5D, and 3D}\label{subsec:modelvars}

Since our dataset contains intrinsically 3D structures, we compared the performances of this architecture using 2D and 3D convolutions.
The former processes individual tomography z-slices (XY plane) independently, and the latter processes several at once (i.e.\ a volume).

We also compared an \say{intermediate} version, which we coined \textit{2.5D}, that processes one tomography z-slice at a time using 2D convolutions, but takes five slices at the input, as if the pairs of data slices above and below were channels of the 2D image.
Table~\ref{tab:modelvars} summarizes these differences.

The visual characteristics of the z-slices are mostly invariant, and we observed a high correlation between adjacent z-slices;
therefore, the 2D and 2.5D models take 2D cuts in the XY plane, as in Fig.~\ref{fig:test-segm-zooms} Fig.~\ref{fig:raw-data}.
\newglossaryentry{tm:zero-order}{
name={ZeroOC},
first={Zero$^{th}$ Order Classifier (ZeroOC)},
description={Classify every voxel with the majority class (matrix).}
}

\newglossaryentry{tm:bin-zero-order}{
name={Bin-\gls{tm:zero-order}},
first={Bin-wise \gls{tm:zero-order}},
description={
Classify a voxel based only on its value.
The majority class of each value is chosen.
This is equivalent to a \gls{tm:zero-order} model per gray level.
}
}


\section{Results}
\label{sec:results}

In this section we present a compilation of qualitative and quantitative results obtained.
The segmentations from the three \gls{modunet} versions presented in Section~\ref{sec:nn} are quantitatively compared, then an ablation analysis and the learning curve of the 2D model are presented.
All the quantitative analyses were made on the test split (see Section~\ref{sec:data}), which contains $1300 \times 1040 \times 300 \approx $ \SI{406e6}{voxels}.
Other images and videos are provided along with further detailed analysis as supplementary material in the~\ref{sec:further-results}.
The trained models and the data used to produce our results are publicly available online~\cite{zenodoModels,zenodoData}.

\paragraph{Qualitative results}

\begin{figure}[!t]

\captionsetup{font=scriptsize, labelfont=scriptsize, width=.9\textwidth}
\centering

\hspace{.01\textwidth}
\begin{subfigure}[b]{.55\textwidth}
\centering
\includegraphics[
height=58mm,
trim=5cm 3cm 1cm 4cm,
clip
]{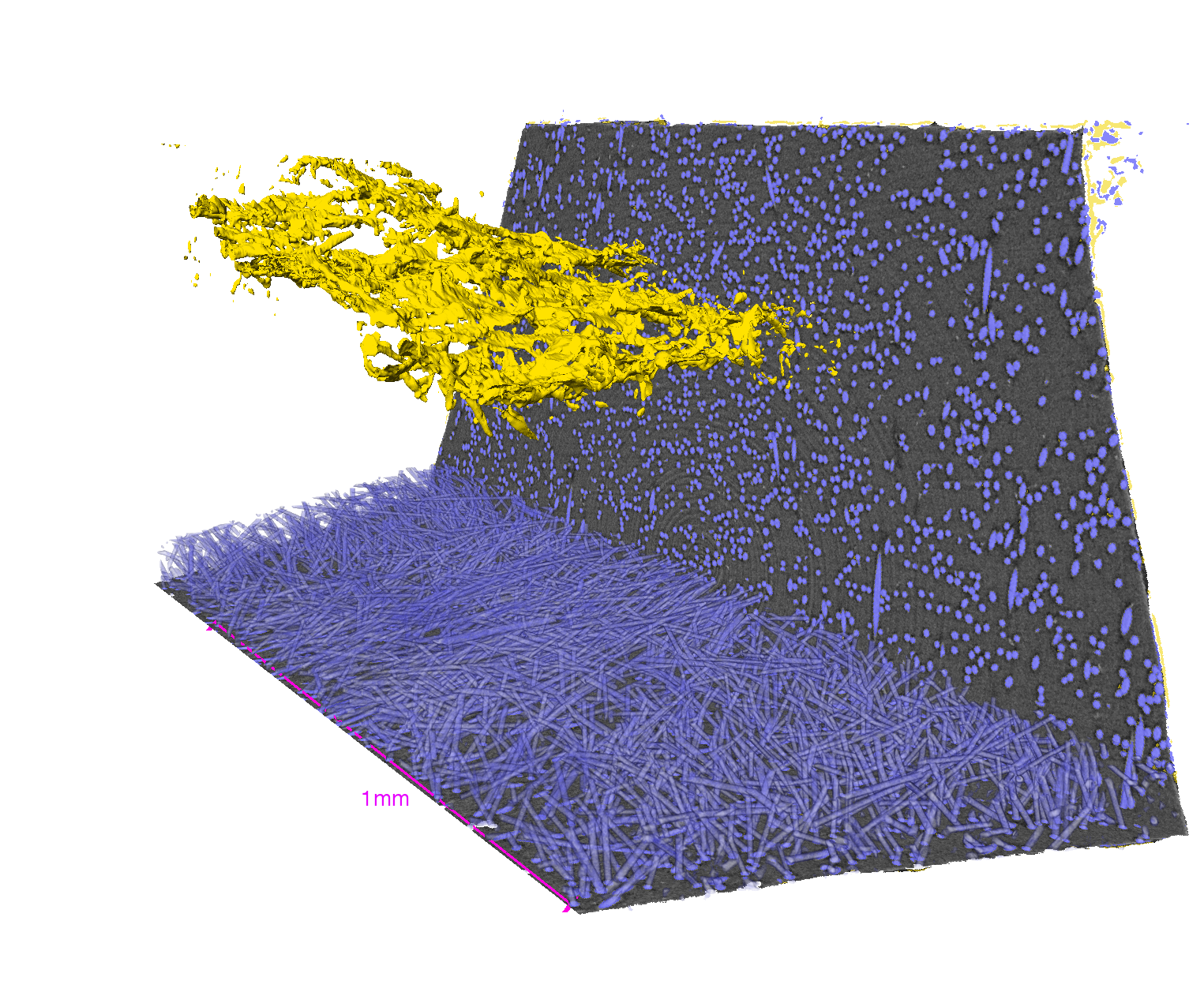}
\caption{}
\label{fig:biax-vol}
\end{subfigure}
\hfill
\begin{subfigure}[b]{.38\textwidth}
\centering
\includegraphics[
height=58mm,
trim=0cm 1cm 0cm 3cm,
clip
]{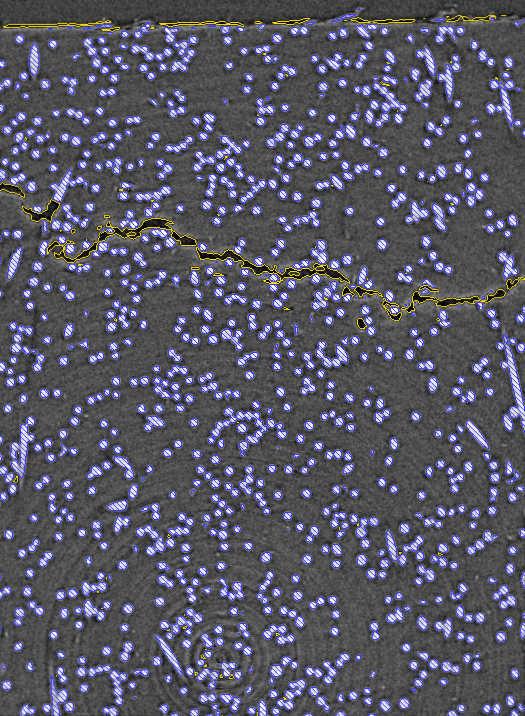}
\caption{}
\label{fig:biax-cut}
\end{subfigure}
\hspace{.01\textwidth}

\caption{
Segmentation of the volume \gls{biax}.
Color code: \textcolor{segm:blue}{blue represents the fiber} and \textcolor{segm:yellow}{yellow represents the porosity}.
Supplementary video: \linkYoutubeBiaxSurface.
(\subref{fig:biax-vol}) Two orthogonal planes inside the specimen;
the fibers are rendered in 3D at the bottom of the volume, and the crack is rendered as a surface.
(\subref{fig:biax-cut}) A crop from the vertical plane in a slice passing through the fracture.
\textcolor{segm:blue}{The fiber segmentation is hatched in blue} and \textcolor{segm:yellow}{the porosity segmentation is contoured in yellow}.
}
\label{fig:biax}

\end{figure}

Figure~\ref{fig:test-segm-zooms} shows two snapshots of the segmentation generated with a 2D model.
Figure~\ref{fig:biax} shows the segmentation obtained with the 2D model from the \gls{biax} volume, used evaluate its usability in terms of processing speed and applicability of the method to other data.
The segmented data was then used to generate a surface of the crack inside it\footnote{\linkYoutubeBiaxSurface} -- we refer to it as the \say{\gls{biax}} volume in the next section.
Using an \gls{nvidia} \gls{qptwok}\footnote{\linkNvidiaQuadroPTwoK} (\SI{5}{\giga B}), it took \SI{32}{minutes} to process a $1579 \times 1845 \times 2002$  ($\approx$ \SI{5800}{Mvoxels}) volume.
This shows that this type of analysis could carried out almost in real time using typical hardware available at a synchrotron beamline.

\paragraph{Baseline}

For the sake of comparison, we considered the expected performance of two theoretical models: the \gls{tm:zero-order} and the \gls{tm:bin-zero-order}.
The \gls{tm:zero-order} relies only on the class imbalance, while \gls{tm:bin-zero-order} takes the individual gray values into consideration, leveraging information from the histograms of each class (Fig.~\ref{fig:data-hist-per-label}) -- see Table~\ref{tab:theoretical-models} for more details.

\paragraph{Quantitative results}

\begin{figure}[!t]

\captionsetup{font=scriptsize, labelfont=scriptsize, width=.9\textwidth}
\centering

\begin{subfigure}[b]{.55\textwidth}

\centering

\includegraphics[height=60mm]{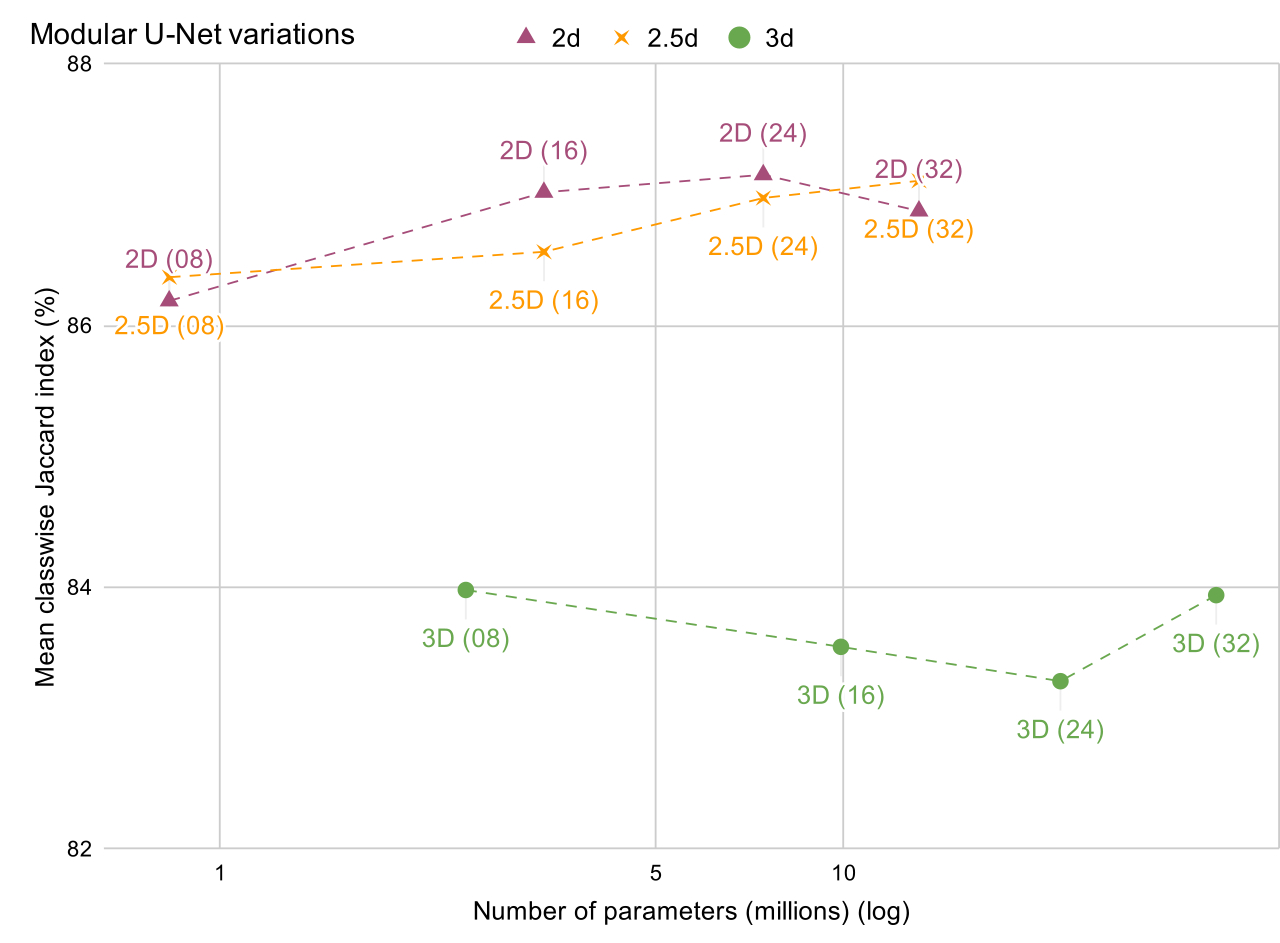}

\captionsetup{font=scriptsize, labelfont=scriptsize, width=.9\textwidth}
\caption{}
\label{fig:models-perf-chart-modelvars}

\end{subfigure}
\hfill
\begin{subfigure}[b]{.43\textwidth}

\centering

\begin{tikzpicture}

\node[inner sep=0] (model_ablation) at (0, 0)
{\includegraphics[height=60mm]{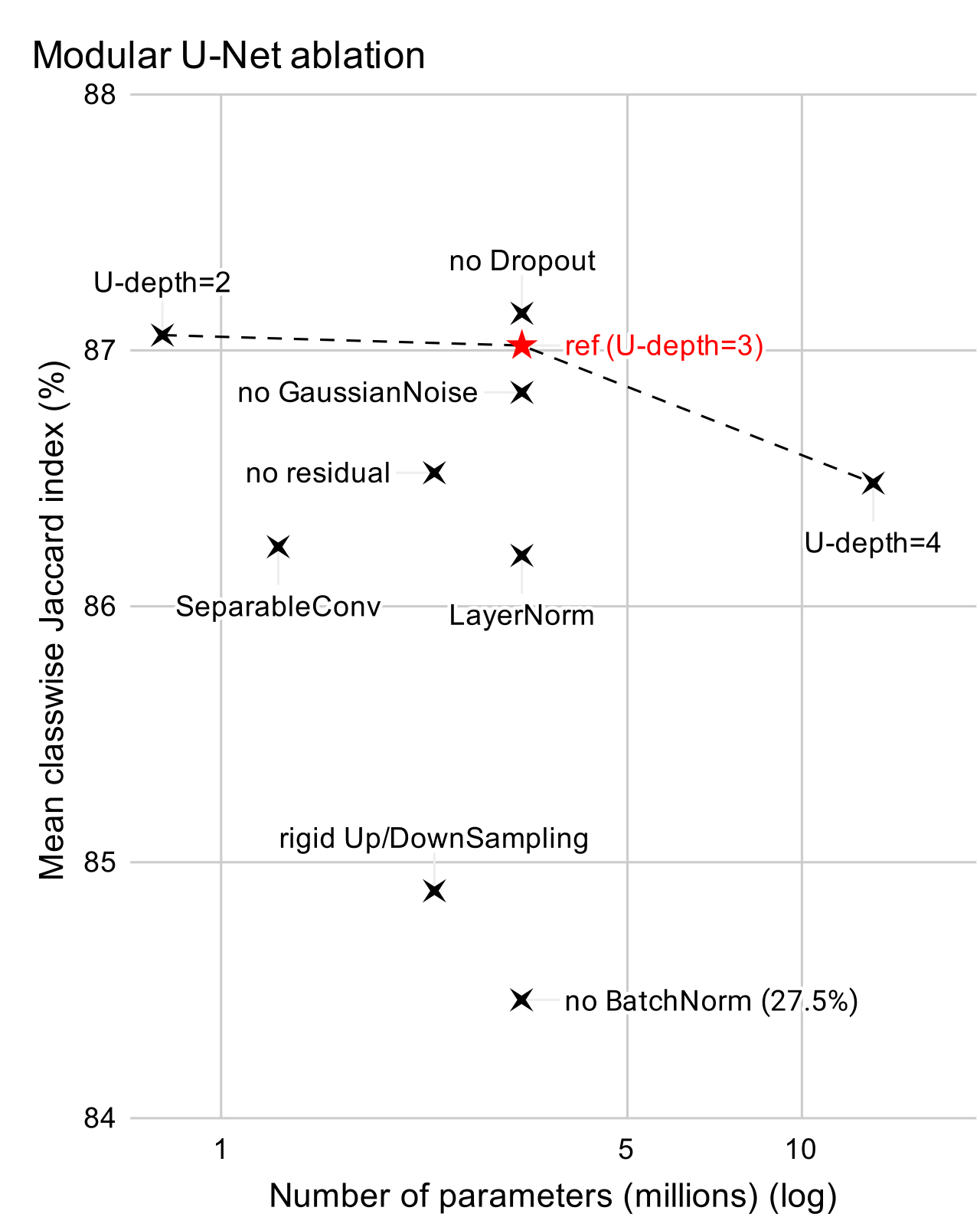}};

\node (nobatch) at (0.15, -1.85) {};
\node (nobatchpointer) at (0.15, -2.50) {};

\draw[->, >=latex] (nobatch) edge (nobatchpointer);

\end{tikzpicture}

\captionsetup{font=scriptsize, labelfont=scriptsize, width=.9\textwidth}
\caption{}
\label{fig:models-perf-chart-modelstripping}

\end{subfigure}

\caption{
\gls{modunet} variations comparison.
On the x-axis, the number of parameters; on the y-axis the mean class-wise \glspl{jaccidx}.
(\subref{fig:models-perf-chart-modelvars}) The \gls{modunet} 2D, 2.5D, and 3D versions are scaled with $ \: \fzero \: $ (in parentheses), the number of filters of the first convolution of the first \ac{convblock}.
(\subref{fig:models-perf-chart-modelstripping}) Components were removed individually, or replaced by alternatives.
Removals: dropout, gaussian noise, residual (skip connection), batch normalization (out of scale).
Replacements: convolutions by separable ones, learnable \ac{downsample}/\ac{upsample} by rigid ones (Fig.~\ref{fig:modunet-blocks}), and \gls{batchnorm} by \gls{layernorm}.
We also compare the effect of the \gls{udepth}, i.e. number of levels in the U structure.
Notice that the data point of \textit{the \gls{batchnorm} removal is out of scale in the y-axis} for the sake of visualization.
}
\label{fig:models-perf-chart}

\end{figure}

Figure~\ref{fig:models-perf-chart} presents a comparison of the three model variations (2D, 2.5D, and 3D) and an ablation study of the 2D model in terms of number of parameters and performance.
The three variations of the \gls{modunet} are evaluated with varying sizes.
The models are scaled with the hyperparameter $ \: \fzero \: $  (values inside the parentheses in Fig.~\ref{fig:models-perf-chart-modelvars}).

The performance is measured using the \gls{jaccidx}, also known as \ac{iou}, on each phase (class).
Our main metric is the arithmetic mean of the three class-wise indices, and the baseline (minimum) is $76.2$\% (Table~\ref{tab:theoretical-models}).
This metric provides a good visibility of the performance differences and resumes the precision-recall trade off;
other classic metrics -- even the area under the \acs{roc}\cite{rocmeaning} curve -- are close to 100\% (see~\ref{sec:further-results}), so the differences are hard to compare.

\paragraph{Ablation study}

Figure~\ref{fig:models-perf-chart-modelstripping} is a component ablation analysis of the 2D model with $ \: \fzero = 16 \: $.
Starting with the 2D model with the default hyperparameters (see Fig.~\ref{fig:modunet-blocks}, and Section~\ref{sec:default-hyperparams}), we retrained other models removing one component at a time.
The learnable up/down-samplings were replaced by \say{rigid} ones (see Fig.~\ref{fig:modunet-blocks}), the 2D convolutions were replaced by separable ones, and the \gls{batchnorm} was replaced by \gls{layernorm}.
Finally, we varied the \gls{udepth} from 2 to 4.

Notice that the model without dropout performed better than the reference model, but we kept it in our default parameters because the same thing did not occur with other variations and sizes.

\paragraph{Learning curve}

\begin{figure}[t]

\captionsetup{font=scriptsize, labelfont=scriptsize, width=\textwidth}
\centering

\includegraphics[height=70mm]{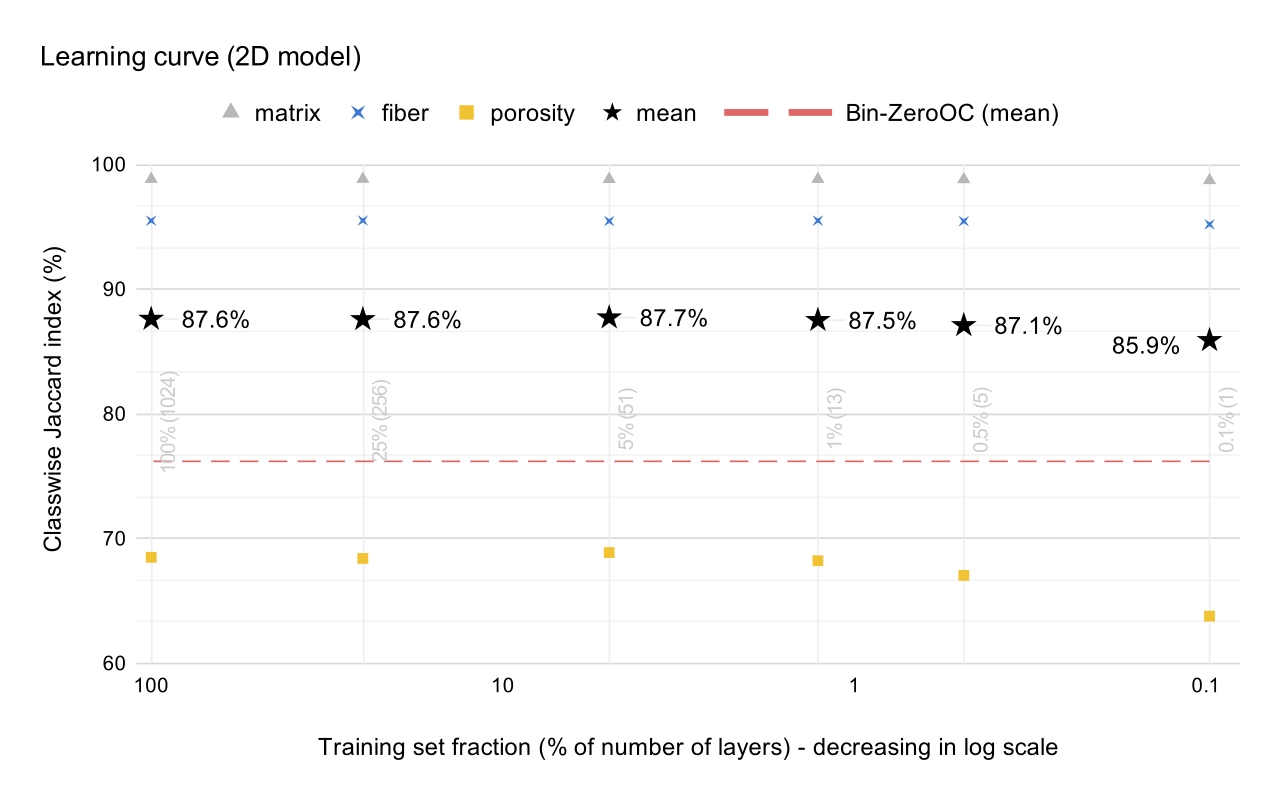}
\caption{
Learning curve of the 2D model with default hyperparameters (\ref{sec:default-hyperparams}).
}
\label{fig:learning-curve}

\end{figure}

Finally, we computed the learning curve of the 2D model (Fig.~\ref{fig:learning-curve}) logarithmically reducing the size of the train dataset from 1024 z-slices until a single layer.

\section{Discussions}\label{sec:discu}
\acresetall

\subsection{Overview}\label{subsec:overview}

Our models, trained in one to three hours\footnote{Up to eight hours for the largest 3D model.}, achieved, qualitatively, very satisfactory results from a Materials Science application point of view, with 87\% of mean class-wise \gls{jaccidx} and an F1-score macro average of 92.4\% (Table~\ref{tab:classif-report}).
We stress the fact that these results were achieved without any strategy to compensate the (heavy) class imbalance (82.3\% of the voxels belong to the class matrix); they may be further improved using, for instance, re-sampling strategies~\cite{pouyanfarDynamicSamplingConvolutional2018, andoDeepOversamplingFramework2017}, class-balanced loss functions~\cite{caoLearningImbalancedDatasets2019, khanStrikingRightBalance2019}, or self-supervised pre-training~\cite{yangRethinkingValueLabels2020}.

The results obtained with another specimen (the \gls{biax} volume), thus with slight variations in the acquisition conditions, were of good quality (inspection by an expert showed no visible error in the segmentation) and way faster than the manual process.
The crack was mostly, and correctly, segmented as porosity without retraining the model, showing its capacity to generalize -- an important feature for its practical use, although some misclassified regions can be seen as holes (missing pieces) in the fracture's surface (Fig.~\ref{fig:biax-vol}).

Moreover, the processing time achieved (\SI{32}{miutes}) is indeed a promising prospect compared to classic approaches.

\subsection{Segmentation errors}\label{subsec:mistakes}

As highlighted in Fig.~\ref{fig:test-segm-zooms}, the model's mistakes (in red), are mostly on the interfaces of the phases, which are fairly comparable to a human annotator's.
We (informally) estimate that they are in the error margin because, in some regions, there is no clear definition of the phases' limits.
The fibers may show smooth, blurred phase boundaries with the matrix, while part of the porosities are under the image resolution.

Another issue is the loss of information (all-zero regions) in some rings (e.g.: Fig.~\ref{fig:raw-data-b}).
In such cases, even though one could deduce that there is indeed a porosity, it is practically impossible to draw a well-defined porosity/matrix interface.

Finally, we reiterate that the ground truth remains slightly imperfect despite our efforts to mitigate these issues.
For instance, in Figure~\ref{fig:test-segm-zooms} we see, inside the C-like shaped porosity, a blue region, meaning that it was \say{correctly} segmented as a fiber -- yet, there is no fiber in it.

\subsection{Model variations}\label{subsec:discussion-model-variations}

Figures~\ref{fig:modelvars-classwise} and~\ref{fig:learning-curve} confirm that, no matter the model variation, the porosity is harder to detect.
Although, the qualitative results are reasonable, and we underline that the \gls{jaccidx} is more sensitive on underrepresented classes because the size of the union will always be smaller (see Equation~\ref{eq:jaccidx}).

Contrarily to our expectations, Figure~\ref{fig:models-perf-chart-modelvars} shows that the 2D model performed systematically better than the 3D (albeit the difference is admittedly small).
We expected the 3D model to perform better because the morphology of the objects in the image are naturally three-dimensional;
besides, other work~\cite{furatMachineLearningTechniques2019} have obtained better results in binary segmentation problems.
We raise two hypotheses about this result: (1) the set of hyperparameters is not optimal, and (2) the performance metric is biased because the annotation process uses a 2D algorithm.

\subsection{Model ablation}\label{subsec:discussion-model-ablation}

Figure~\ref{fig:models-perf-chart-modelstripping} contains a few interesting findings about the hyperparameters of the \gls{modunet}:

\begin{enumerate}

\item \label{itm:updown-sampling} using learnable up/down-sampling operations indeed gives more flexibility to the model, improving its performance compared to \say{rigid} (not learnable) operations;

\item \label{itm:sep-conv} separable convolutions slightly hurt the performance, but it reduces the number of parameters by 60\%;

\item \label{itm:dec-depth} decreasing the \gls{udepth}, therefore shrinking the receptive field, improved the performance while reducing $75\%$ of the model size;
on the other hand, increasing the depth had the opposite effect, multiplying the model size by four, while degrading the performance;

\item \label{itm:batchnorm} \gls{batchnorm} is essential for the training -- notice that the version without \gls{batchnorm} is out of scale in Fig.~\ref{fig:models-perf-chart-modelstripping}, and its performance corresponds to the \gls{tm:zero-order} model (Table~\ref{tab:theoretical-models});

\end{enumerate}

\paragraph{Model depth (item~\ref{itm:dec-depth})}

This finding gives a valuable information for our future work because using shallower models require less memory (i.e.: bigger crops can be processed at once), making it possible to accelerate the processing time.
We hypothesize that the necessary receptive field is smaller than the depth-three model's.
Therefore, a spatially bigger input captures irrelevant, spurious context to the classification.

\subsection{Learning curve}\label{subsec:discussion-learning-curve}

Figure~\ref{fig:learning-curve} highlights the most promising finding in our results.
Our model was capable of learning with only 1\% of the training dataset (about ten z-slices) even with no \say{smart} strategy to select the layers in the experiment (the training volume was sequentially reduced along the z-axis), and even a single layer was sufficient to achieve nearly the same performance.


\section{Conclusion}
\label{sec:conclusion}

An annotated dataset of a three-phase composite material was presented, and a reinterpretation of the \gls{unet} architecture as a modularized structure was proposed as a solution to scale up the segmentation of such images.
Our models achieved satisfactory results showing a promising venue to automate processing pipelines of \acsp{xct} of this material with only a few annotated tomography slices.
The \gls{modunet} is a conceptually more compact interpretation of its precursor, providing a more abstract representation of this family of network architectures.

2D and 3D versions of the \gls{modunet} were compared, showing that both were capable of learning the patterns in a human-made annotation, but the former was systematically better than the latter.
An ablation study provided insights about the hyperparameters of our architecture, especially revealing that we might further accelerate the processing with smaller models.
We also qualitatively analyzed the usability of our models on an image of a different specimen, confirming that it is a viable solution.

Our code is available on \gls{gh}\footnote{\linkTts}, and the data and trained models referred here are publicly available on Zenodo~\cite{zenodoData,zenodoModels}.


\section{Future work}
\label{sec:future-work}

Carrying on the encouraging results found in this work, we envision scaling up the use of our models to process other volumes and 4D \acp{xct}.
We also plan identifying better strategies to deal with ill-defined regions (e.g.: matrix/fiber blurred interfaces), an issue modestly mitigated in our work, to improve our approach's usability.
Two possibilities are considered: (1) post-processing the class probabilities to detect bad predictions, for instance using the method proposed by \cite{energyBasedOutOfDistributionDetection};
(2) define a special class for uncertain regions.
Finally, we might as well search for better ways to measure a prediction's consistency with respect to the objects' 3D morphology.


\section{Acknowledgement}\label{sec:acknowledgement}

João P C Bertoldo would like to acknowledge funding from the \gls{bigmeca} research initiative\footnote{\linkBigmeca}, funded by Safran and \gls{mines}.
Henry Proudhon would like to acknowledge beam time allocation at the \gls{psyche} beamline of \gls{soleil}\footnote{\linkSoleil} for the proposal number 20150371.

\bibliography{ref}

\newpage
\appendix


\section{Data: further details}
\label{sec:data-further}

\begin{figure}[!h]

\captionsetup{font=scriptsize, labelfont=scriptsize, width=.9\textwidth}
\centering

\begin{subfigure}[b]{.55\textwidth}
\centering

\includegraphics[height=58mm]{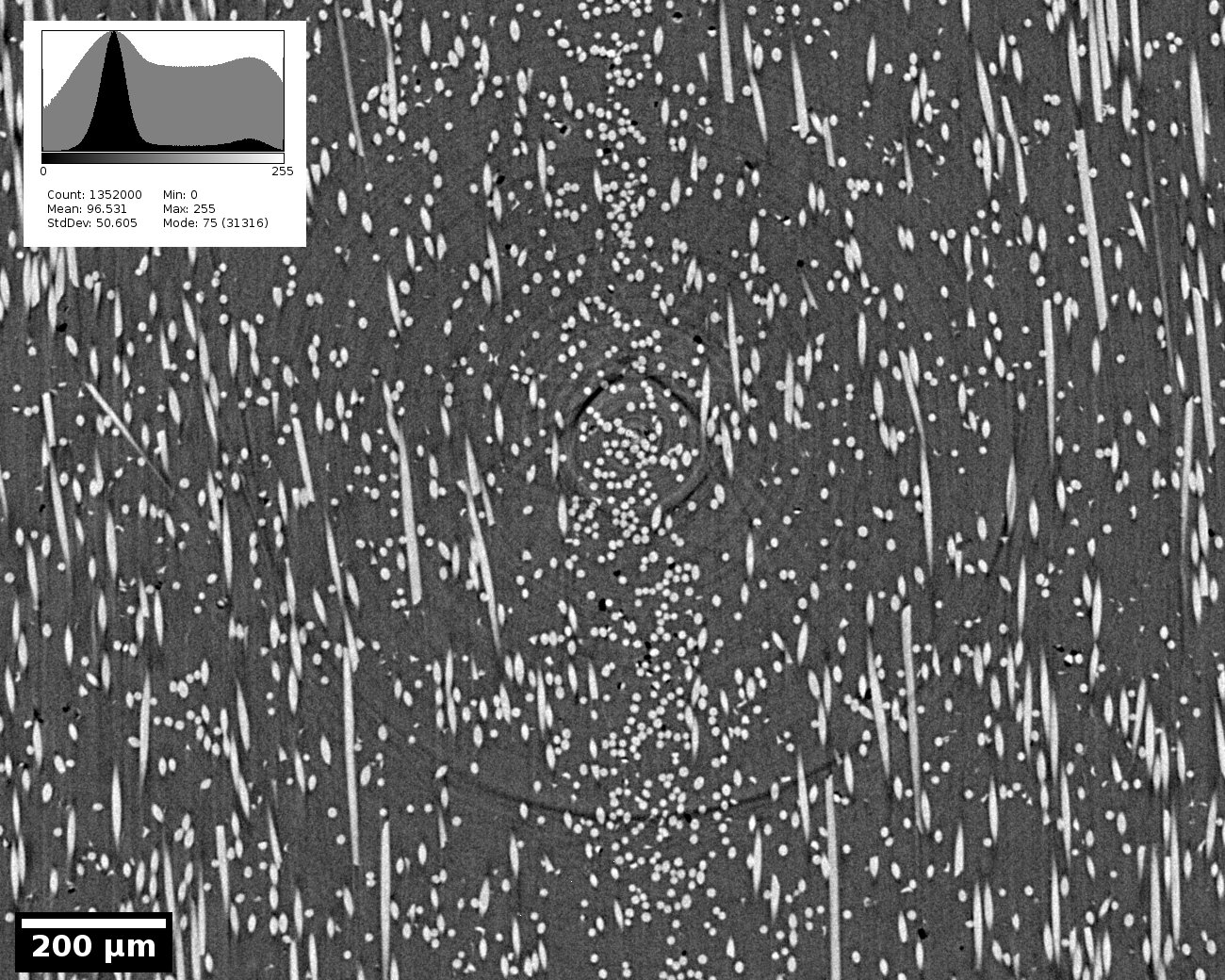}

\captionsetup{font=scriptsize, labelfont=scriptsize, width=.9\textwidth}
\caption{
A $1300 \times 1040$ slice on the XY plane of the volume \gls{train-val-test}.
On the upper left corner, a histogram of the gray level values in the image;
linear scale in black, log scale in gray.
}
\label{fig:raw-data-a}

\end{subfigure}
\hfill
\begin{subfigure}[b]{.44\textwidth}
\centering

\includegraphics[height=58mm]{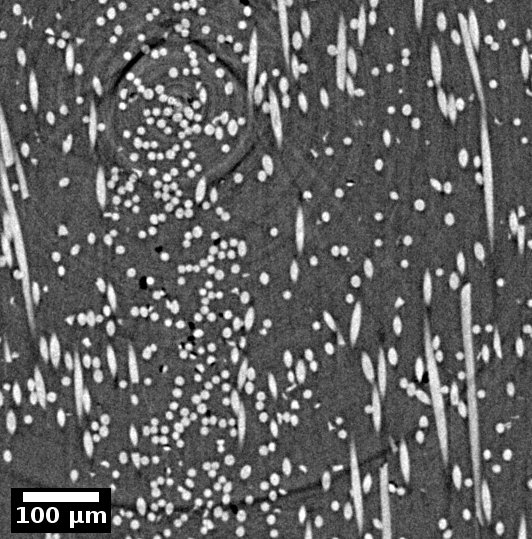}

\captionsetup{font=scriptsize, labelfont=scriptsize, width=.9\textwidth}
\caption{
Zoom. Ring artifacts, from the acquisition process, can be as dark as porosities, making it harder to segment such regions.
}
\label{fig:raw-data-b}

\end{subfigure}

\caption{
\Gls{gf-poly} raw tomography.
Supplementary video: \linkYoutubeRawData.
}
\label{fig:raw-data}

\end{figure}

\subsection{Data annotation}
\label{subsec:data-annotation-further}

We annotated the data in two phases.
First, the fiber and the porosity were separated from the matrix independently using \acl{srg}~\cite{srg}.
As the results carried a considerable amount of artifacts in the porosity phase, we manually corrected them with a second procedure.

\paragraph{Data annotation step 1: \acl{srg}~\cite{srg}}
To generate the seeds, we applied contrast and brightness transformations to enhance the information on the phase of interest, then applied \gls{non-local-means}~\cite{nonlocalmeans1,nonlocalmeans2} to attenuate the ring artifacts, and finally used manual thresholds to select \say{easy} voxels (mostly, regions without class superposition in Fig.~\ref{fig:data-hist-per-label}).
Finally, we run it on each tomography z-slice independently.

We observed that some regions were poorly segmented due to the ring artifacts, creating false, larger porosities.
To mitigate this issue, we manually corrected part of the artifacts, reducing their size while keeping it consistent with adjacent z-slices.

\paragraph{Data annotation step 2: artifacts correction}

We separated 2D blobs (connected components) in the porosity phase, then extracted region properties (e.g.area, aspect ratio, etc) to find outliers -- mostly, porosities significantly larger than the average.
Using \gls{avizo}, we then corrected imperfections manually editing the annotation.
Most problematic regions were on pronounced rings, where it is hard to define the borders of a porosity (e.g.: Fig.~\ref{fig:raw-data}), so we cleaned ill-defined porosities conservatively shrinking their volume/borders based on the layers above and below.

\subsection{Ground truth analysis}
\label{subsec:gt-analysis}

\begin{figure}[!ht]

\captionsetup{font=scriptsize, labelfont=scriptsize, width=.9\textwidth}
\centering

\includegraphics[width=.7\textwidth]{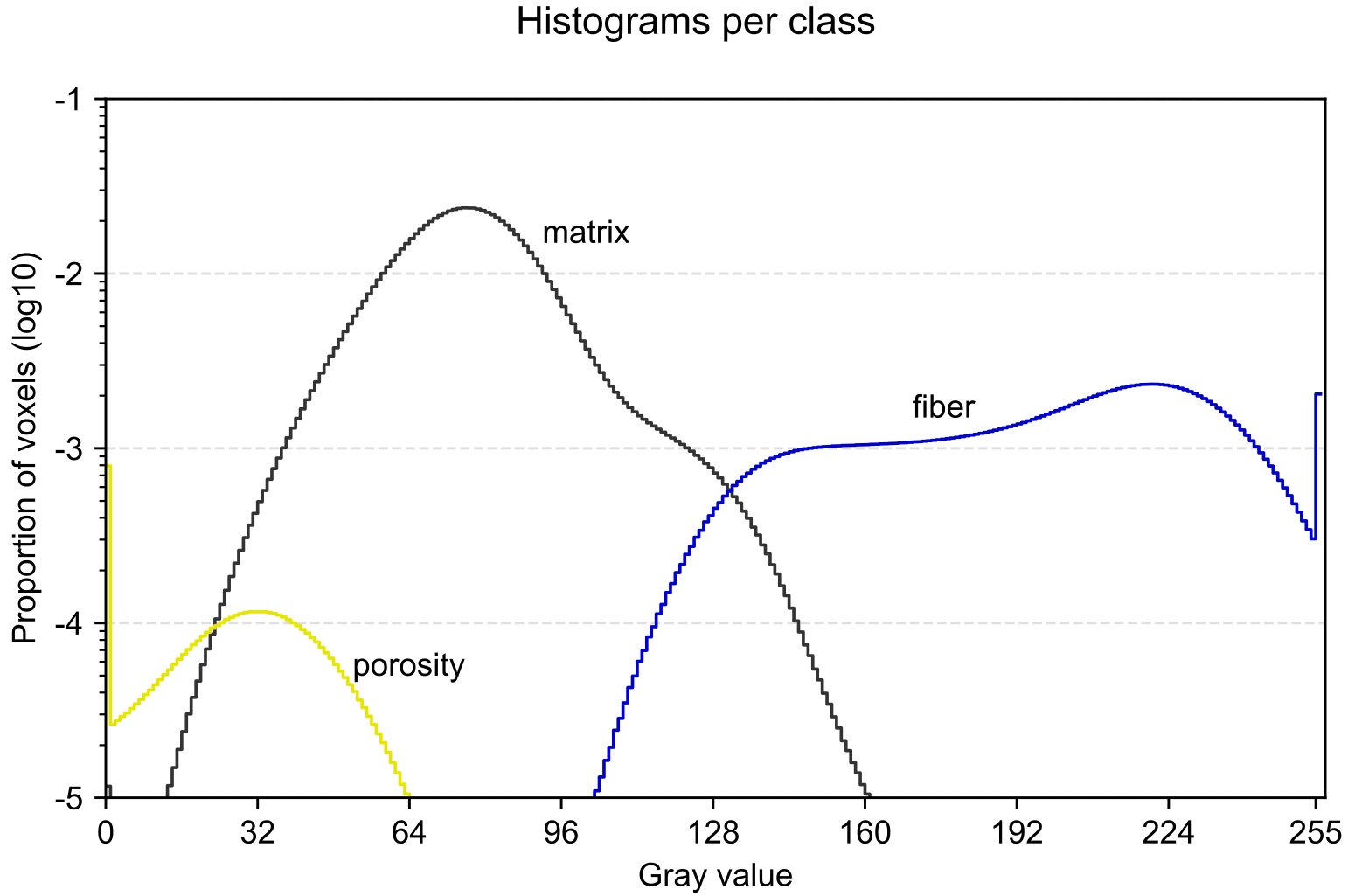}

\caption{
Glass fiber-reinforced \gls{poly} gray value (normalized) histograms (one per class).
The histogram is normalized globally, i.e. a bin's value is the proportion of voxels out of all the voxels (all classes confounded).
The superposition of the classes' value ranges make it impossible to segment the image with a threshold on the gray values.
}

\label{fig:data-hist-per-label}

\end{figure}

\paragraph{Class-wise histogram}
Figure~\ref{fig:data-hist-per-label} shows a normalized gray level histogram
per class (the normalization is relative to all the classes confounded).
Using a threshold approach is naturally prone to imprecise results because a voxel's gray value is insufficient to determine its class.
This illustrates how this volume cannot be segmented using a threshold.
An example can be seen in Fig.~\ref{fig:raw-data-b}, where the rings are as dark as the porosities.

\subsection{Data availability}
\label{subsec:data-availability}

The volumes mentioned here (Table~\ref{tab:public-volumes}) are available on \gls{zenodo}~\cite{zenodo}.
Notice that the volumes (\ref{vol:data}) and (\ref{vol:gt}) contain all the three splits (train, validation, test) together (1900 z-slices), while the volumes (\ref{vol:test-segm}) and (\ref{vol:test-err}) correspond to their last 300 z-slices.
The values in the segmentation volumes (predictions and ground truth) are 0, 1, and 2, which respectively correspond to the phases matrix, fiber, and porosity.
The volumes (\ref{vol:crack-data}) and (\ref{vol:crack-pred}) correspond to the volume in Fig.~\ref{fig:biax}.
The \code{.raw} files have complementary \code{.raw.info} files containing metadata (volume dimensions and data type) about its respective volume.

\begin{table}[!ht]

\centering
\tiny
\renewcommand{\arraystretch}{1.7}
\captionsetup{font=scriptsize, labelfont=scriptsize, width=0.9\textwidth}
\caption{
Published 3D volumes: all the data necessary to train and test the models presented in this paper are publicly available on Zenodo~\cite{zenodoData}.
A demo of how to read the data is available on \linkOpenVolumesDemo{\gls{gh}}.
}
\label{tab:public-volumes}
\begin{tabular}{ccp{65mm}}
\textbf{.zip file}              & \textbf{.raw file}          & \multicolumn{1}{c}{\textbf{Description} }                                                                                                 \\
\hline
\multirow{2}{*}{pa66.zip}       & \code{pa66.raw} \vollabel{vol:data}                    & Data (gray level image stack) of the \gls{train-val-test} volume.                                                     \\
\cline{2-3}
& \texttt{pa66.ground\_truth.raw} \vollabel{vol:gt}      & Ground truth segmentation of the \gls{train-val-test} volume.                                                          \\
\hline
\multirow{2}{*}{pa66\_test.zip} & \code{pa66.test.prediction.raw} \vollabel{vol:test-segm}    & Segmentation generated by the best 2D model on the \textit{test }set.                                                                     \\
\cline{2-3}
& \texttt{pa66.test.error\_volume.raw} \vollabel{vol:test-err} & Disagreement between the ground truth and the model's prediction on the \textit{test} set: 1 means \textit{incorrect}, 0 means \textit{correct}.   \\
\hline
\multirow{2}{*}{crack.zip}      & \texttt{crack.raw} \vollabel{vol:crack-data}                 & Data of the non-annotated volume containing a crack inside.                                                                      \\
\cline{2-3}
& \texttt{crack.prediction.raw} \vollabel{vol:crack-pred}       & Segmentation generated with the best 2D model on the crack volume.
\end{tabular}

\end{table}

Figure~\ref{fig:raw-data} was generated in \gls{fiji}\cite{fiji} with the volume (\ref{vol:data}).
Figure~\ref{fig:test-segm-zooms} was generated in \gls{avizo} with volumes (\ref{vol:data}) and (\ref{vol:test-err}), which is derived from volumes (\ref{vol:gt}) and (\ref{vol:test-segm}).
Figure~\ref{fig:biax-cut} was generated in \gls{avizo} with volumes (\ref{vol:crack-data}) and (\ref{vol:crack-pred}).
All the supplementary videos were generated in \gls{avizo}.

\newpage


\section{Default hyperparameters}\label{sec:default-hyperparams}

Parameters not mentioned are the default in TensorFlow 2.2.0.

\begin{table}[!h]

\centering
\footnotesize

\captionsetup{font=scriptsize, labelfont=scriptsize, width=0.9\textwidth}
\caption{
\gls{modunet} variations: input, convolutional layer, and output nature (2D or 3D).
}
\label{tab:modelvars}

\begin{tabular}{r|ccc}
\textbf{Model} & \textbf{Input (data)} & \textbf{Convolution} & \textbf{Output (segm.)} \\
\hline
\Tstrut
\textbf{2D}    & 2D                                                                              & 2D                   & 2D                                                                                \\
\textbf{2.5D}  & 3D                                                                                & 2D                   & 2D                                                                                \\
\textbf{3D}    & 3D                                                                                 & 3D                   & 3D
\end{tabular}
\end{table}

\begin{table}[!h]

\centering
\footnotesize

\captionsetup{font=scriptsize, labelfont=scriptsize, width=0.9\textwidth}
\caption{
Default hyperparameters.
}
\label{tab:default-hyperparams}

\def\arraystretch{1.5}

\begin{tabular}{cccc}
\textbf{Parameter}                                                                       & \textbf{2D}   & \textbf{2.5D}                                                       & \textbf{3D}          \\ \hline
U-depth                                                                                  & 3             & 3                                                                   & 3                    \\ \hline
Convolution kernel                                                                       & 3 $\times$ 3     & 3 $\times$ 3                                                           & 3 $\times$ 3 $\times$ 3    \\ \hline
Batch size                                                                               & 10            & 10                                                                  & 10                   \\ \hline
Crop shape                                                                               & 160 $\times$ 160 & 160 $\times$ 160 $\times$ 5 & 32 $\times$ 32 $\times$ 32 \\ \hline
Dropout                                                                                  & 10\%          & 10\%                                                                & 10\%                 \\ \hline
\begin{tabular}[c]{@{}c@{}}Gaussian noise (zero mean) \\ standard deviation\end{tabular} & 0.03          & 0.03                                                                & 0.03                 \\ \hline
f0                                                                                       & 16            & 16                                                                  & 16                   \\ \hline
\begin{tabular}[c]{@{}c@{}}Up/Down-sampling stride\\ or Max pooling size\end{tabular}    & 2 $\times$ 2     & 2 $\times$ 2                                                           & 2 $\times$ 2 $\times$ 2    \\ \hline
\multicolumn{1}{l}{BatchNorm momentum}                                                   & 0.5           & 0.5                                                                 & 0.5
\end{tabular}
\end{table}
\newpage


\section{Training}\label{sec:training}

\paragraph{Loss function}

We trained our models using a custom loss inspired on the \gls{jaccidx}, also known as \ac{iou}.

\begin{definition}\label{def:jaccidx}

Let $ \: A \: $ and $ \: B \: $ be two discrete sets.
The \gls{jaccidx} $ \: J \in \interval{0}{1} \: $ is

\begin{equation}
\label{eq:jaccidx}
J(A, B) = \frac{|A \cap B|}{|A \cup B|} = \frac{|A \cap B|}{|A| + |B| - |A \cap B|}
\end{equation}

\end{definition}

We adapt, similarly to~\cite{powerJaccardLosses}, the second form in Equation~\ref{eq:jaccidx} to define the multi-class \gls{jacc2} loss for a batch of voxels (a batch of 2D or 3D images unraveled on the spatial dimensions) as follows.

\begin{definition}\label{def:jacc2}

Let $ \: \gtvox{i} \in \{0,1\}^{C} \: $ be the one-hot-encoding ground truth vector of the voxel at position $ \: i \in \sequenceset{B} \: $ in a batch of $ \: B \: $ voxels $ \: \gtvol \in \{0,1\}^{B \times C} \: $, where $ \: \gtvox{ic} \in \{0, 1\} \: $ is its value in the $ \: c\text{-th} \: $ position.

\begin{equation*}
\gtvox{ic} =
\begin{cases}
1, & \text{if the voxel $i$ belongs to the class} \; c \in \segvoxdom \\
0 & \text{otherwise}
\end{cases}
\end{equation*}

A model's last activation map, a per-voxel softmax, is a tensor $ \: \probavol \in \interval{0}{1}^{B \times C} \: $, where each row is a probability vector $ \: \probavox{i} \in \interval{0}{1}^C \: $, and the component $ \: \probavox{ic}$ corresponds to the probability assigned to the class $ \: c \: $.

The \gls{jacc2} loss $ \: J2 \in \interval{0}{1} \: $ of the batch $ \: (\gtvol, \probavol) \: $ is

\begin{align}
\label{eq:jacc2}
J2(\gtvol, \probavol) &=
1 - \frac{
\sum_{i = 1}^{N} \sum_{c = 1}^{C} \gtvox{ic} \probavox{ic}
}{
\sum_{i = 1}^{N} \sum_{c = 1}^{C} \gtvox{ic} \gtvox{ic}
+ \sum_{i = 1}^{N} \sum_{c = 1}^{C} \probavox{ic} \probavox{ic}
- \sum_{i = 1}^{N} \sum_{c = 1}^{C} \gtvox{ic} \probavox{ic}
} \\
&= 1 - \frac{
\sum_{i = 1}^{N} \probavox{i*}
}{
N + \sum_{i = 1}^{N} \left( \left(\sum_{c = 1}^{C} \probavox{ic}^2 \right) - \probavox{i*} \right)
}
\end{align}

where $ \: \probavox{i*} = \sum_{c = 1}^{C} \gtvox{ic} \probavox{ic}  \: $ is the probability assigned to the correct class of the voxel $ \: i $.

\end{definition}

Notice that the $ \: J2 \in \interval{0}{1}$, which is convenient because it can be expressed as a percentage.
$ \: J2 = 100\% \: $ is a completely uncorrelated estimation, and $ \: J2 = 0\% \: $ is a perfect replication of the ground truth.

\paragraph{Optimizer and \acl{lr}}

We used \gls{adabelief}~\citep{adabelief}, an optimizer that combines the training stability and fast convergence of adaptive optimizers (e.g.: \gls{adam}~\cite{adam}) and good generalization capabilities of accelerated schemes (e.g.: \ac{sgd}~\cite{sgd}).
We used a learning rate of $10^{-3}$ for 100 epochs (10 batches each with batch size 10), followed by a linearly decaying rate until $10^{-4}$ in another 100 epochs.
Adam gave equivalent results but took longer (more epochs) to converge.

\paragraph{Data augmentation}

In order to increase the variability of the data, random crops are selected from the data volume, then a random geometric transformation (flip, \ang{90} rotation, transposition, etc) is applied.
As our training dataset is reasonably large, we used a simple data augmentation scheme, but richer transformations may be applied as long as the transformations result in credible samples.

\paragraph{Implementation and hardware}

We trained our models using \gls{keras}~\cite{keras} with \gls{tf}'s~\cite{tensorflow2015} \acs{gpu}-enabled version\footnote{\href{\linkPypiTfGpu}{\gls{tf-gpu} on \gls{pypi}}} with \gls{cuda} $10.1$ running on two \gls{nvidia} \gls{qpfourk}\footnote{\linkNvidiaQuadroPFourK} (2x \SI{8}{\giga B}).
The implementation of our experiments is available on \gls{gh}\footnote{\linkTts}.
\newpage


\section{Further results}
\label{sec:further-results}

In this section we present complementary performance metrics, images, and videos from the segmentation generated by the 2D model (see Section~\ref{sec:nn}) using our default hyperparameters (\ref{sec:default-hyperparams}) on the test split.
Videos are available as supplementary material\footnote{\linkYoutubeTestVolSegm, \linkYoutubeTestVolErr, \linkYoutubeTestVolSegmTwo}.

Table~\ref{tab:theoretical-models} presents the expected performances of the two baseline models we considered (Section~\ref{sec:results}).

\begin{table}[!h]

\scriptsize
\captionsetup{font=scriptsize, labelfont=scriptsize, width=.95\textwidth}
\centering

\caption{
Expected performance of baseline theoretical models in terms of class-wise \gls{jaccidx} (\%).
}
\label{tab:theoretical-models}

\begin{tabular}{cp{45mm}cccc}

\textbf{Model}      & \textbf{Description} & \multicolumn{1}{l}{\textbf{Matrix }} & \multicolumn{1}{l}{\textbf{Fiber }} & \multicolumn{1}{l}{\textbf{Porosity }} & \multicolumn{1}{l}{\textbf{Mean }} \Bstrut \\

\hline

\gls{tm:zero-order}     & {\scriptsize \glsdesc{tm:zero-order}    } & 81.0                                     & 0                                       & 0                                          & 27.0      \Tstrut \\

\gls{tm:bin-zero-order} & {\scriptsize \glsdesc{tm:bin-zero-order}} & 98.4                                        & 94.2                                       & 35.9                                          & 76.2  \Tstrut

\end{tabular}
\end{table}

Table~\ref{tab:classif-report} shows a report with classic classification metrics and the \gls{jaccidx} by class along with their macro/micro averages.
As one can see, the accuracy, precision, and recall of the matrix and the fiber are all close to 100\%, while the porosity's scores are considerably lower.

\begin{table}[!ht]
\centering

\captionsetup{font=scriptsize, labelfont=scriptsize, width=10cm}
\caption{
2D \gls{modunet} ($ \: \fzero = 16 \: $) classification report.
}
\label{tab:classif-report}

\scriptsize

\begin{tabular}{ccccccc}
\textbf{class}       & \textbf{accuracy (\%)} & \textbf{precision (\%)} & \textbf{recall (\%)} & \textbf{f1 (\%)} & \textbf{jaccard (\%)} & \textbf{support}   \\
\hline
\textbf{matrix}      & 99.0                   & 99.3                    & 99.4                 & 99.4             & 98.8                  & 334.4 million      \\
\textbf{fiber}       & 99.2                   & 97.7                    & 97.6                 & 97.6             & 95.3                  & 69.1 million       \\
\textbf{porosity}    & 99.8                   & 84.2                    & 76.5                 & 80.2             & 66.9                  & 2.1 million        \\
\hline
\textbf{macro avg.}  & 99.3                   & 93.7                    & 91.2                 & 92.4             & 87.0                  & -                  \\
\textbf{micro avg.}  & 99.0                   & 99.0                    & 99.0                 & 99.0             & -                     & -
\end{tabular}
\end{table}

\begin{figure}[!ht]

\captionsetup{font=scriptsize, labelfont=scriptsize, width=.9\textwidth}
\centering

\includegraphics[height=60mm]{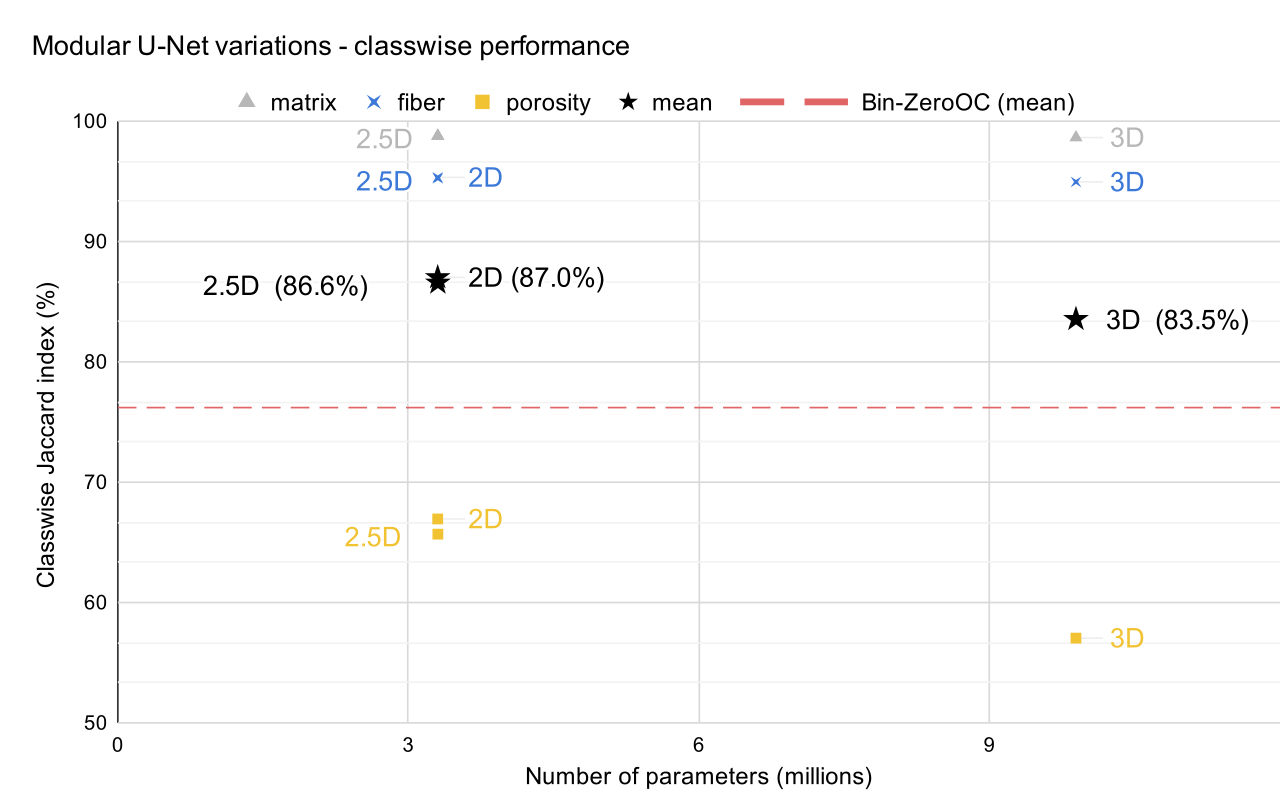}

\caption{Class-wise \gls{jaccidx} \gls{modunet} 2D, 2.5D, and 2D variations.}
\label{fig:modelvars-classwise}

\end{figure}

Figure~\ref{fig:modelvars-classwise} shows that, as the matrix and the fiber phases have high scores, the mean \gls{jaccidx} is driven by the performance on the porosity detection, which can also seen in Fig.~\ref{fig:learning-curve}.
Recall that the baseline model \gls{tm:bin-zero-order} is expected to have a \gls{jaccidx} of 35\% on the porosity and 76.2\% of mean (Table~\ref{tab:theoretical-models}).

\begin{figure}[!ht]

\captionsetup{font=scriptsize, labelfont=scriptsize, width=.9\textwidth}
\centering

\includegraphics[width=.6\textwidth]{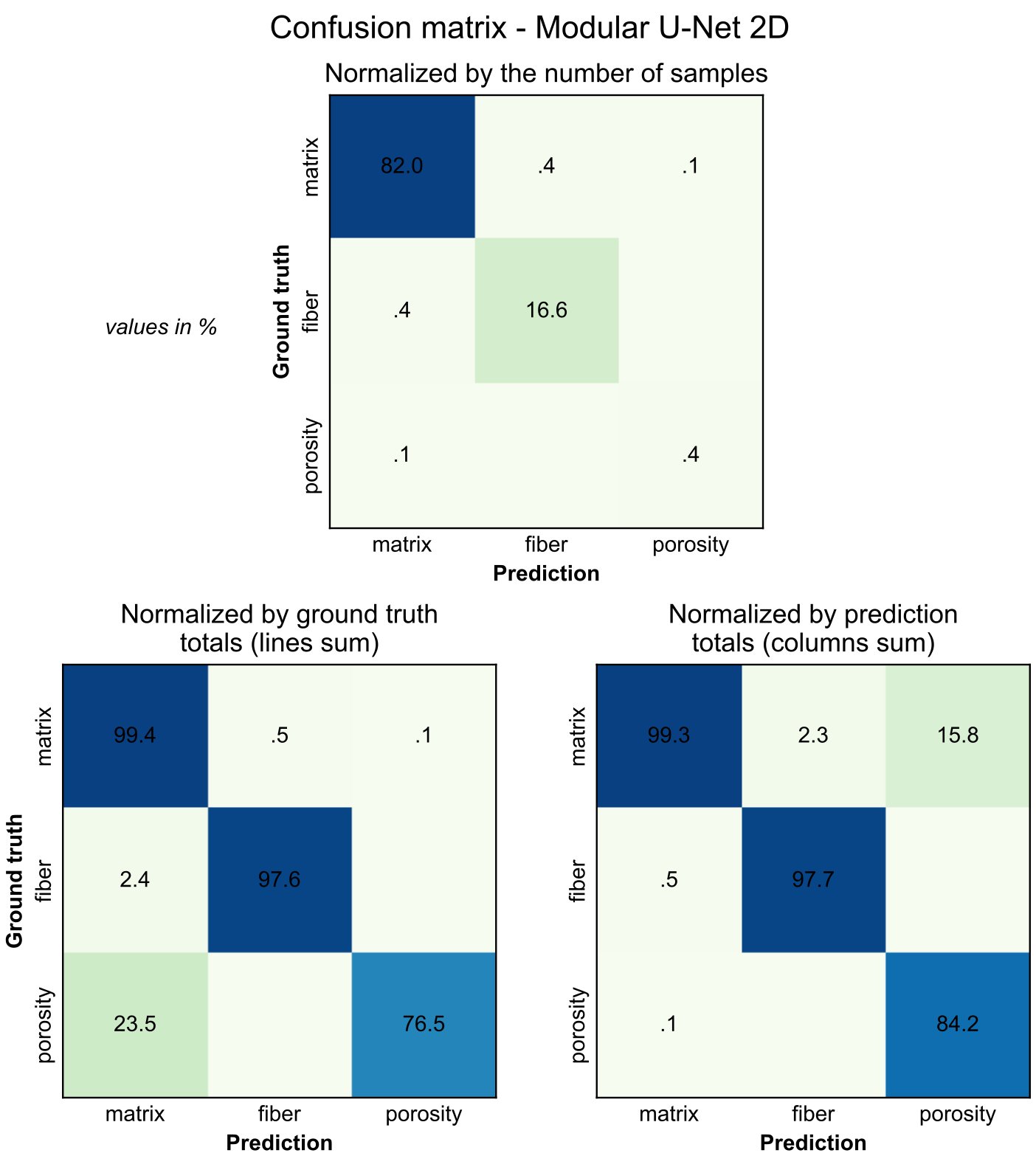}

\caption{
Confusion matrices of the 2D model on the test set normalized in three different ways.
(top) Normalized by the sum of all cells confounded.
(bottom left) Normalized by the sum of \textbf{ground truth} labels (a.k.a. support) of each class (each \textbf{line} sums up to 100\%); the diagonal corresponds to the recall values.
(bottom right) Normalized by the sum of \textbf{predicted} labels of each class (each \textbf{column} sums up to 100\%); the diagonal corresponds to the precision values.
}

\label{fig:confusion-matrix}

\end{figure}

Figure~\ref{fig:confusion-matrix} presents detailed confusion matrices.
The top one is expressed percentage of the number of voxel counts.
The two bottom ones are normalized, respectively, by row and column, and their diagonals correspond to the recall and the precision of each class.

\end{document}